\documentclass[%
 reprint,
 amsmath,amssymb,
 aps,
prb,
]{revtex4-2}
\usepackage{graphicx}
\usepackage{dcolumn}
\usepackage{bm}
\usepackage{textcomp} 
\usepackage[utf8]{inputenc}
\usepackage{newunicodechar}
\newunicodechar{−}{$-$}
\usepackage{textgreek}
\begin{document}
\title{\textbf{Deterministic Helicity Locking of Bloch Skyrmions in Centrosymmetric Systems}} 
\author{Jayaseelan Dhakshinamoorthy }
 \email{djn@niser.ac.in}
\author{Hitesh Chhabra}%
\author{Ajaya K Nayak}%
 \email{ajaya@niser.ac.in}
\affiliation{%
School of Physical Sciences, National Institute of Science Education and Research, An OCC of Homi Bhabha National Institute, Jatni 752050, India\\
}%

\date{\today}

\begin{abstract}
Magnetic skyrmions in centrosymmetric materials exhibit Bloch-type spin textures with degenerate helicity states due to the absence of Dzyaloshinskii–Moriya interaction (DMI), resulting in random nucleation and uncontrolled chirality. Here, we present a comprehensive micromagnetic study demonstrating a fully DMI-free strategy for deterministic helicity control by interfacing ferromagnetic (FM) stripes or notch structures with centrosymmetric magnetic (CM) films. We first show that a geometrically constrained configuration—comprising two FM stripes with opposite in-plane magnetizations—stabilizes skyrmions with a selected helicity, either clockwise (CW) or counterclockwise (CCW). We further extend this concept to achieve deterministic nucleation of CW or CCW skyrmions using current pulses applied to an FM notch patterned on the CM film. The combined effects of the FM stripe/notch geometry and interfacial exchange coupling generate dipolar fields that lift the helicity degeneracy, enabling controlled formation of skyrmions with fixed chirality. These results establish FM/CM heterostructures as a robust, DMI-free platform for deterministic generation and guided motion of helicity-locked skyrmions, opening new pathways for advanced spintronic applications.
\end{abstract}
\maketitle


\section{INTRODUCTION}

Magnetic skyrmions are vortex$-$like swirling spin textures with topologically protected configurations, first experimentally observed in the chiral magnet MnSi \cite{muhlbauer_skyrmion_2009} and subsequently found in several non-centrosymmetric/centrosymmetric bulk magnets \cite{yu_near_2011, tokunaga_new_2015,yu_variation_2017,nayak_magnetic_2017,kurumaji_skyrmion_2019} and heavy-metal/ferromagnet heterostructure thin films \cite{caretta_fast_2018,hsu_electric-field-driven_2017,montoya_transport_2022,legrand_room-temperature_2020,soumyanarayanan_tunable_2017,ahrens_skyrmions_2023,heigl_dipolar-stabilized_2021}.
Due to their nanoscale size, stability, and unique dynamical properties, skyrmions have emerged as promising candidates for next-generation spintronic technologies \cite{kang_skyrmion-electronics_2016,fert_magnetic_2017,wiesendanger_nanoscale_2016}, including race track memory and logic gates \cite{mankalale_skylogicproposal_2019,chen_sky-ram_2019,chauwin_skyrmion_2019,zhu_control_2024, raab_brownian_2022}. 
In the case of non-centrosymmetric magnets, breaking of the inversion symmetry gives rise to an asymmetric Dzyaloshinskii–Moriya interaction (DMI), which locally tilts the uniform magnetic state, giving them a fixed helicity.
In contrast, the competing long-range dipolar energy and uniaxial magnetic anisotropy (UMA) are primarily responsible for the stabilization of skyrmions in centrosymmetric magnets \cite{kurumaji_skyrmion_2019,heinze_spontaneous_2011,yu_magnetic_2012,hayami_effective_2017,ozawa_zero-field_2017}, although frustrated magnetic exchange and higher order exchanges are also responsible for stabilizing small skyrmions \cite{kurumaji_skyrmion_2019,paddison_magnetic_2022,paul_role_2020}. 
However, skyrmions found in centrosymmetric systems, where the presence of inversion symmetry does not support the DMI, exhibit degenerate energy with respect to their helicity states   \cite{kong_diverse_2024,hou_currentinduced_2023}.
However, from a topological point of view, these skyrmions are equivalent to their DMI-stabilized counterparts \cite{hou_observation_2017,wei_currentcontrolled_2021}.

In systems where skyrmions are stabilized by the competition between dipolar interactions and uniaxial magnetic anisotropy, the underlying mechanism can be described by the Hamiltonian interaction for a pair of spins as \cite{sarkar_magnetic_2025},
\begin{equation}
	\begin{aligned}
		H = & -\frac{\mu_0 \, \gamma_i \gamma_j \hbar^2}{4\pi |\mathbf{r}|^3}
		\left[ 3(\mathbf{S}_i \cdot \hat{\mathbf{r}})(\mathbf{S}_j \cdot \hat{\mathbf{r}}) - \mathbf{S}_i \cdot \mathbf{S}_j \right] \\
		& - \frac{1}{2} \sum_{i \neq j} J_{ij} \, (\mathbf{S}_i \cdot \mathbf{S}_j)
		- K_u \sum_i (\mathbf{S}_i \cdot \hat{\mathbf{z}})^2\\
		& - \sum_i (\mathbf{B}_{\mathrm{ext}} \cdot \mathbf{S}_i),
	\end{aligned}
\end{equation}
where the first term represents the dipole–dipole interaction, the second term corresponds to the Heisenberg exchange, the third term denotes the uniaxial anisotropy, and the final term accounts for the Zeeman interaction with an external magnetic field $B_{ext}$. 
$\mathbf{S}_i$ and $\mathbf{S}_j$ denote the spins at the $i_{th}$ and $j_{th}$ atomic sites, $J_{ij}$ represent the exchange coupling strength, $\gamma_i$ and $\gamma_j$ are the gyromagnetic ratios,  $\hat{\mathbf{r}}$ is a unit vector connecting the two spins, $|\mathbf{r}|^3$ is their separation distance, and $K_u$ is the uniaxial anisotropy constant. 
The skyrmions are characterized by helicity, which defines the rotation sense (clockwise (CW) or counterclockwise (CCW)) of the in-plane component of the magnetization \cite{yu_variation_2017}. 
For example, most centrosymmetric magnets host Bloch-type skyrmions with both clock-wise (CW) and counter clock-wise (CCW) helicities, i.e. $-\pi/2$ and $+\pi/2$,  respectively  \cite{yu_magnetic_2012,hirschberger_skyrmion_2019,li_magnetic_2023}. 
Although this helicity freedom may be advantageous for some applications \textbf{ \cite {Zhou_2025,Petrovi_2025}}, deterministic control of helicity is crucial for spintronics applications where helicity states can be used as a classical or quantum bit \cite{nagaosa_topological_2013,xia_universal_2023}. 
Helicity locking enables magnetic textures to maintain a well-defined chirality even in the absence of DMI, thereby stabilizing the spin configuration against external perturbations. 
This robustness is critical to achieve reliable information storage and signal transmission \cite{lin_ginzburg-landau_2016}. 
Overall, it provides a pathway to utilize skyrmions in CM materials that are easier to fabricate and integrate into existing device architectures. 
In DMI-based skyrmion systems, helicity is inherently fixed by the sign and direction of the DMI vector and can only be modified by tuning the DM vector, for example, by spin–orbit coupling \cite{shibata_towards_2013,karube_controlling_2018,liu_modulation_2024}.
In contrast, helicity in DMI-free centrosymmetric materials can be modulated via external perturbations, such as electric or magnetic fields \cite{yao_controlling_2020,yao_vector_2022}, strong stray dipolar fields \cite{navas_route_2019}, or nano-structure \cite{sun_creating_2013,verba_overcoming_2018}. 
Although few approaches have been proposed to manipulate the helicity, deterministic control of helicity in centrosymmetric magnetic heterostructures remains largely unexplored and is essential for advancing skyrmion-based spintronics.

In this work, we propose a mechanism to control the helicity of Bloch-type skyrmions through dipolar coupling by introducing  ferromagnetic (FM) stripes/notches with in-plane (IP) anisotropy to a UMA based centrosymmetric ferromagnet (CM),as illustrated in Fig. 1a. The degeneracy between skyrmions with opposite helicities is lifted by a tunable dipolar field arising from exchange coupling between the CM layer and adjacent FM layers, enabling deterministic helicity switching. Furthermore, by locally modifying the FM anisotropy from IP to out-of-plane (OP) at the nucleation site and applying an external current, we demonstrate controlled generation of skyrmions with selected helicities, including paired CW–CCW skyrmions as well as random-helicity states, as shown in Fig. 1(b). This strategy provides a pathway toward helicity locking and programmable skyrmion creation in centrosymmetric magnetic thin films, advancing their integration into spintronic devices.

\section{COMPUTATIONAL METHOD}
To validate our analytical approach and gain deeper insight into helicity locking in the proposed systems, we have performed micromagnetic simulations using the MuMax\textsuperscript{3 } solver \cite{vansteenkiste_design_2014}. The computational mesh was discretized with a cell size of 2 × 2 × 1 nm³. For the CM layer the material parameters used are : saturation magnetization M\textsubscript{s} = 1.6 × 10\textsuperscript{6} A/m, exchange constant A\textsubscript{ex} = 10 pJ, uniaxial anisotropy K\textsubscript{u} = 15 × 10\textsuperscript{5} J/m\textsuperscript{3 }oriented along the z-axis, and Gilbert damping constant $\alpha$ = 0.9, chosen to accelerate relaxation \cite{vansteenkiste_design_2014}. The FM strips are modeled with the same M\textsubscript{s} and A\textsubscript{ex} values as the CM layer, but with a larger uniaxial anisotropy of 30 × 10\textsuperscript{5} J/m\textsuperscript{3 } aligned along the in-plane direction ($\pm x$), to emulate the rigid magnetization of the FM layers. The thickness of the FM stripes is fixed at 2 nm. In contrast, strip width and inter-strip separation are systematically varied to understand their effect on skyrmion stabilization. The interlayer exchange coupling between the CM and FM layers was scaled by a factor ($A_\mathrm{ex}^*$) of 0.5. 
For deterministic helicity control under applied in-plane currents (J\textsubscript{x}, J\textsubscript{y}), the spin-transfer torque (STT) generated by the CM layer nucleates skyrmions with locked helicity. In the simulations, FM notches are modeled with the same M\textsubscript{s} and A\textsubscript{ex} as the CM layer, but with a reduced uniaxial anisotropy of 8 × 10\textsuperscript{5} J/m\textsuperscript{3 } oriented along both the in-plane ($\pm x$) and out-of-plane directions. Micromagnetic simulations were performed using the Landau–Lifshitz–Gilbert (LLG) equation augmented with adiabatic and non-adiabatic STT terms, following the Zhang–Li model \cite{hou_currentinduced_2023,albert_effect_2020,bernstein_spin-torque_2025}. The system was initialized with random magnetization in the CM layer and uniform magnetization in the FM strip. A perpendicular field of 120 mT was applied to reach equilibrium, after which spin-polarized current pulses with density J = (x, y, 0) and polarization P = 0.6 were introduced to drive the nucleation and motion of the skyrmion. The non-adiabaticity parameter was set to $\beta$ = 0.05, consistent with the reported values \cite{bernstein_spin-torque_2025}. Minor adjustments in M\textsubscript{s,} and K\textsubscript{u} were made to ensure numerical stability and reproduce realistic FM behavior. Detailed simulation procedures are provided in the \textbf{Supporting information \cite{SM}}. Mesh-convergence tests verified that the results were free of discretization artifacts. 

\begin{figure}[tb!]
\centering
\includegraphics[width=1\linewidth]{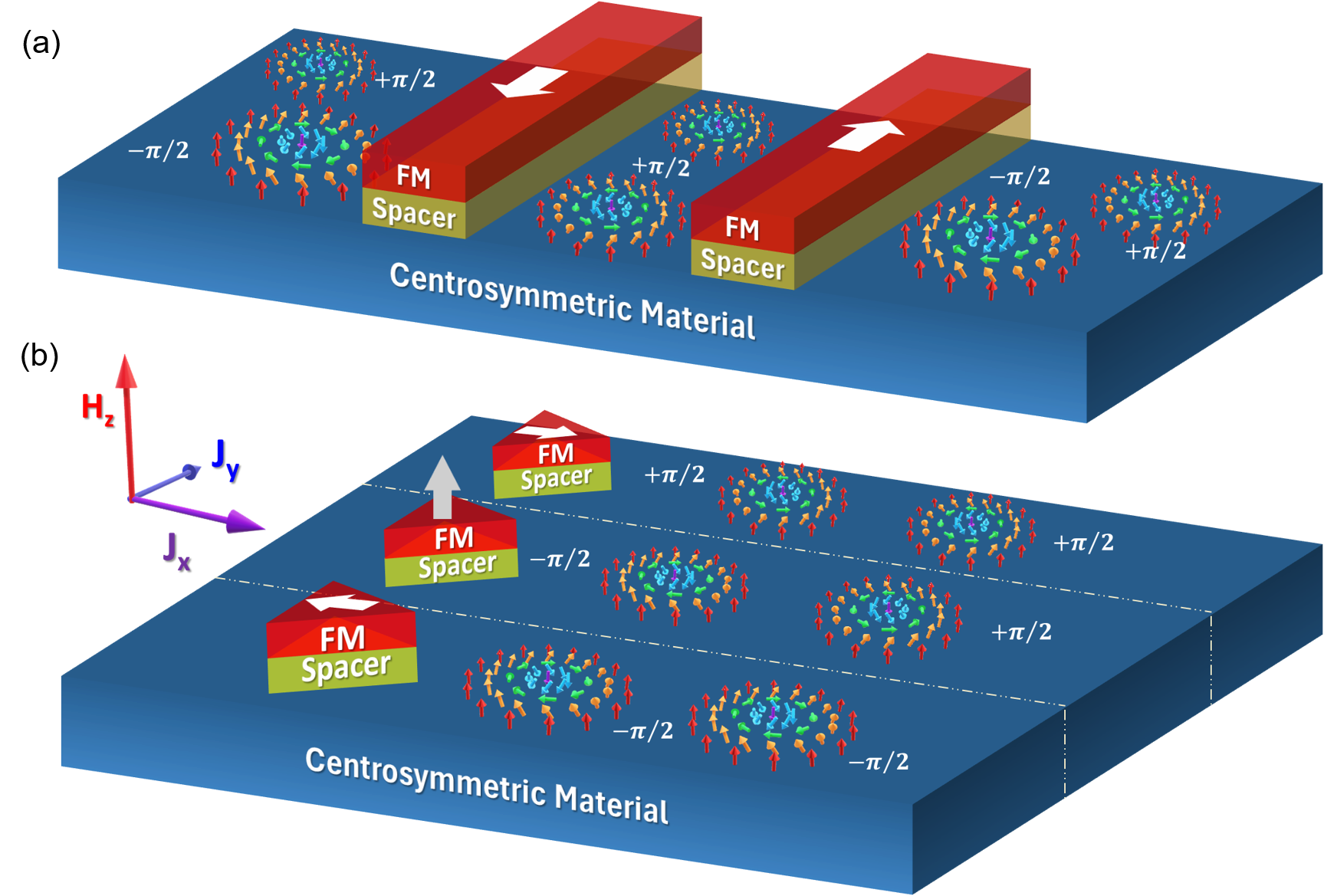} 
\caption{\label{fig:FIG1} Design scheme for helicity locking in a centrosymmetric magnetic thin film system. (a) Schematic illustration showing ferromagnetic (FM)/spacer stripes with opposite in-plane (IP) magnetization on a centrosymmetric ferromagnetic thin film. Opposite skyrmion helicities ($\pm\pi/2$) are stabilized outside the stripe region, while only counterclockwise (CCW, $\eta$ =  $+\pi/2$) skyrmions are stabilized between the stripes. (b) Deterministic skyrmion nucleation with locked helicity using triangular FM notch with IP and OP magnetic moments and under applied in-plane currents (J\textsubscript{x}, J\textsubscript{y}). The FM layer with OP ($+ z$) magnetization nucleates skyrmions with both helicities, whereas the in-plane ($\pm x$) FM layer nucleates only CCW ($\eta$ =  $+\pi/2$) or CW ($\eta$ =  $-\pi/2$) skyrmions.}
\end{figure}


\begin{figure*}[t]
    \centering \includegraphics[width=0.8\textwidth,keepaspectratio]{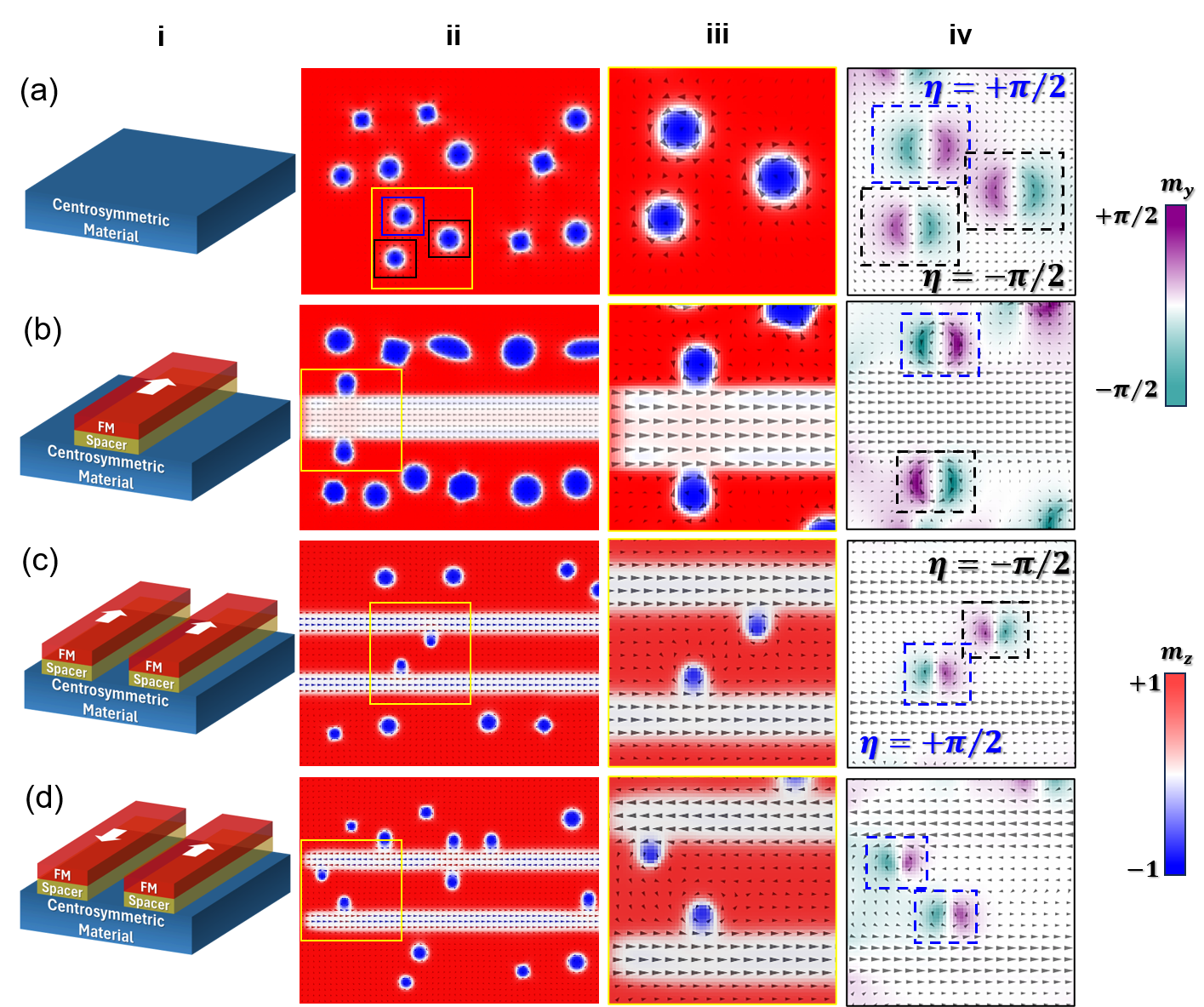} 
    \caption{\label{fig:widepng2} Micromagnetic simulations of dipolar-stabilized skyrmions in a heterostructures with different geometry. (a) Schematic of the simulated geometry along with their corresponding magnetization (m\textsubscript{y}, m\textsubscript{z}) map  for (a) a standalone CM film, (b) CM film with a single FM/spacer layer, (c) Bi-FM/spacer with parallel dipole configuration, (d) Bi-FM/spacer with antiparallel dipole configuration. For each simulated geometry, the schematic, m\textsubscript{z} magnetic profile, zoomed views of the m\textsubscript{z} magnetic profile shown in a boxed, and the with the m\textsubscript{y} magnetization profile are shown under the column i, ii, iii, and iv, respectively. Color bars indicate
     the magnetization components (m\textsubscript{y}, m\textsubscript{z}) in the equilibrium state. }
\end{figure*}

\section{RESULTS AND DISCUSSIONS}

\subsection{FM-Stripe-Mediated Helicity Locking}

\begin{figure*}[!t]
	\centering \includegraphics[width=0.8\textwidth,keepaspectratio]{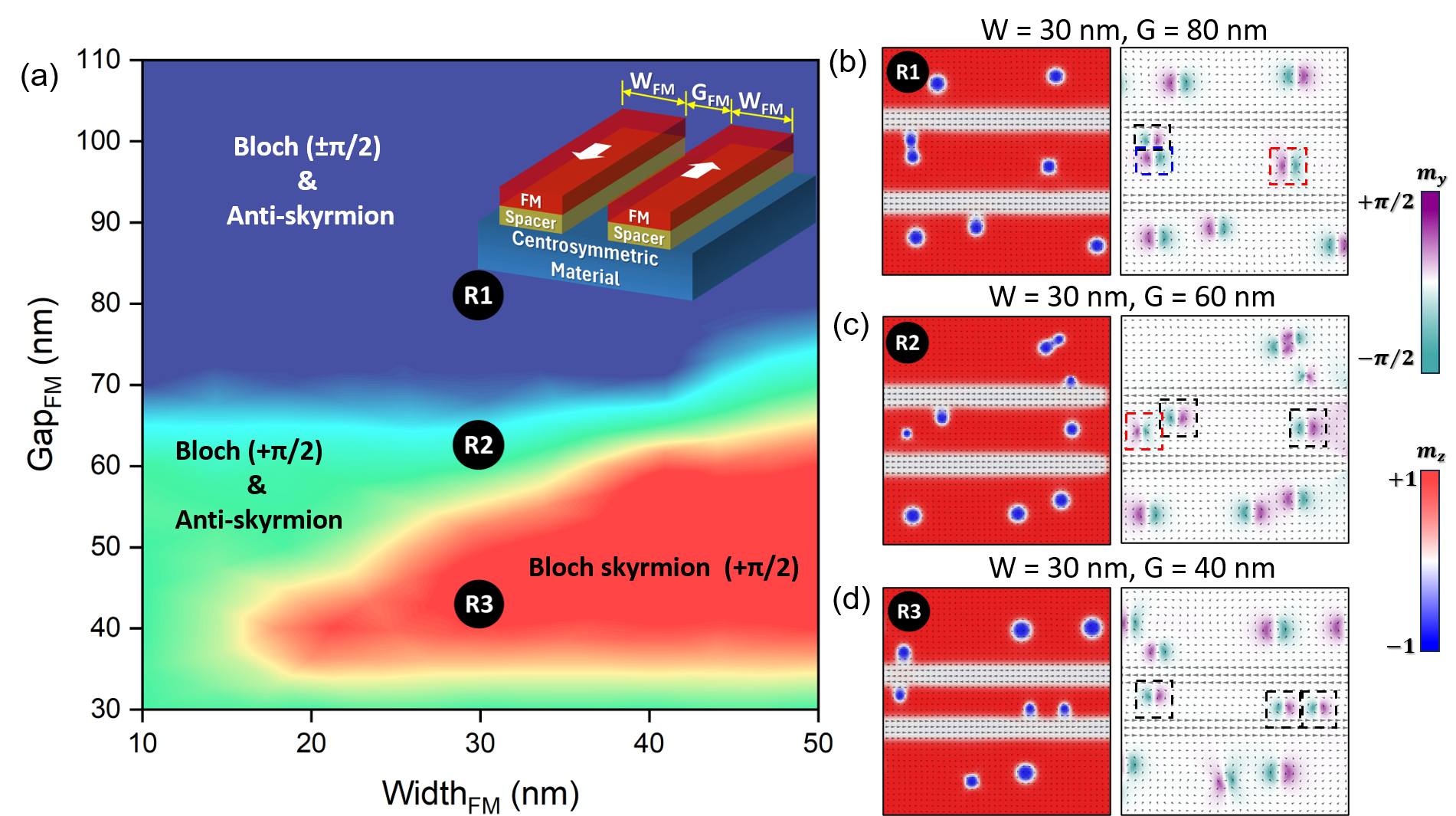} 
	\caption{\label{fig:FIG3}(a) Micromagnetic phase diagram of FM-coupled centrosymmetric film heterostructures as a function of FM width (W\textsubscript{FM}) and gap (G\textsubscript{FM}), Inset: schematic of the FM/Spacer/CM heterostructure with variable W\textsubscript{FM} and G\textsubscript{FM}. (b–d) Representative equilibrium spin textures corresponding to points (R1-R3) in (a): OP magnetization maps (m\textsubscript{z}, red = spin-up, blue = spin-down) and helicity-resolved in-plane magnetization maps. (color scale encodes (m\textsubscript{y}, m\textsubscript{z}), arrows denote local magnetization). ‘The map "m\textsubscript{y}" highlights individual cores, illustrating the transition from helicity-degenerate Bloch and anti-skyrmion states (R1), to coexisting helicity-locked Bloch skyrmion states ($\eta$ = $+\pi/2$)/ anti-skyrmion states (R2), and finally to fully helicity-locked Bloch skyrmion states ($\eta$ = $+\pi/2$) (R3). Bloch skyrmions with helicities $\eta$ = $+\pi/2$ and $-\pi/2$, and anti-skyrmions are highlighted by black, blue, and red squares, respectively.
	}
\end{figure*}


We start with a random magnetization state of the CM thin film with OP anisotropy and then relaxed the system to an equilibrium magnetic configuration, as shown in Fig. 2a.
The m\textsubscript{z} magnetization profile of the simulated spin configuration show the stabilization of isolated skyrmions with the coexistence of CW ($\eta$ = $-\pi/2$) and CCW ($\eta$ = $+\pi/2$) helicities. 
The presence of degenerate helicity states is clearly evident from the zoomed view of the m\textsubscript{z} and  m\textsubscript{y} magnetization profile.
When a single FM stripe with in-plane magnetic anisotropy is placed on top of the CM layer, the effective interlayer exchange field couples the local magnetization variation of the CM layer to that of the FM one, as shown in Fig. 2b. 
In this scenario, the skyrmion helicities within the coupled region are no longer energetically degenerate, but influenced by the  FM stripe. 
This is evident from  the simulated  m\textsubscript{z} and m\textsubscript{y} magnetization profiles, where the skyrmions close to the top and bottom of the IP FM stripe exhibit $+\pi/2$ and $-\pi/2$ helicities, respectively, as seen in Fig. 2b. 
The above helicity selectivity led us to place two FM stripes separated by a small gap and having parallel in-plane magnetization on top of the CM layer. It is found that the skyrmions formed in the gap region display $+\pi/2$ helicity near the bottom FM stripe and $-\pi/2$ helicity near the top one (Fig. 2c). 
We utilize the above concept for the deterministic stabilization of a single helicity by placing two FM layers with antiparallel alignment of the IP magnetic moment, as depicted in Fig. 2d.
This configuration enforces uniform locking of the skyrmion helicities across the coupled region with a single  helicity ($\eta$ = $+\pi/2$, CCW). 
The zoomed image and the ‘\textit{m\textsubscript{y}}’ configuration clearly show that the IP magnetic moments of both the skyrmions are in the same configuration, confirming the helicity locking in the system. 
Overall, these results demonstrate that while helicity is degenerate in isolated centrosymmetric films, coupling to FM structures provides a deterministic means to select, lock, and spatially organize skyrmion helicities without requiring any interfacial DMI.


To further study the role of IP magnetized FM stripes on helicity locking, we systematically vary the width (W\textsubscript{FM}) and inter-strip gap (G\textsubscript{FM}) for the antiparallel FM strips configuration shown in {Fig. 2d.} 
As mentioned earlier, the addition of FM strips introduces stray-fields coupling that progressively lifts the helicity degeneracy, depending on the FM layer geometry. 
For a large G\textsubscript{FM} and narrow W\textsubscript{FM}, the stray-field interaction is weak, and hence helicity-degenerate Bloch skyrmions coexist with anti-skyrmions (Fig. 3a-R1). 
Reducing G\textsubscript{FM } and increasing W\textsubscript{FM} enhances the dipolar-field interaction, leading to partial helicity locking where Bloch skyrmions with $\eta$ = $+\pi/2$ coexist with anti-skyrmions (Fig. 3a-R2). 
At sufficiently small G\textsubscript{FM} and large W\textsubscript{FM}, a strong dipolar coupling completely removes the degeneracy, stabilizing only helicity-locked Bloch skyrmions with $\eta$ = $+\pi/2$ (Fig. 3a-R3). 
The magnetization maps shown in Figures 3b-d for different regions of Fig. 3a confirm the effect of W\textsubscript{FM} and G\textsubscript{FM} on the stabilization of different types of magnetic textures. 
These results demonstrate that dipolar fields from FM structures provide a robust mechanism to enforce deterministic skyrmion helicity in centrosymmetric systems, offering a symmetry-independent alternative to DMI for chirality control. 
This approach complements recent experimental reports on skyrmion stabilization in stray field-coupled soft magnetic materials \cite{verba_helicity_2020,bunyaev_controlling_2025}. 
Like the dipolar field of the FM stripes, the interfacial exchange coupling between the FM stripes and the CM layer also plays an important role for the helicity locking of skyrmions in the present system [Supplement].
At weak coupling ($A_\mathrm{ex}^*$ $\ll$ 0.1), insufficient dipolar interaction from the FM layer leads to the coexistence of Bloch skyrmions with both helicities ($\pm \pi/2$), and anti-skyrmions occasionally appear. Increasing $A_\mathrm{ex}^*$ improves the interlayer exchange, progressively stabilizing a single helicity state ($\eta$ = $+\pi/2$) and eliminating opposite-helicity and anti-skyrmion configurations.

\subsection{Deterministic Helicity-Locking via Skyrmion Nucleation at FM/CM Notch}

\begin{figure*}[!t]
\centering \includegraphics[width=0.8\textwidth,keepaspectratio] {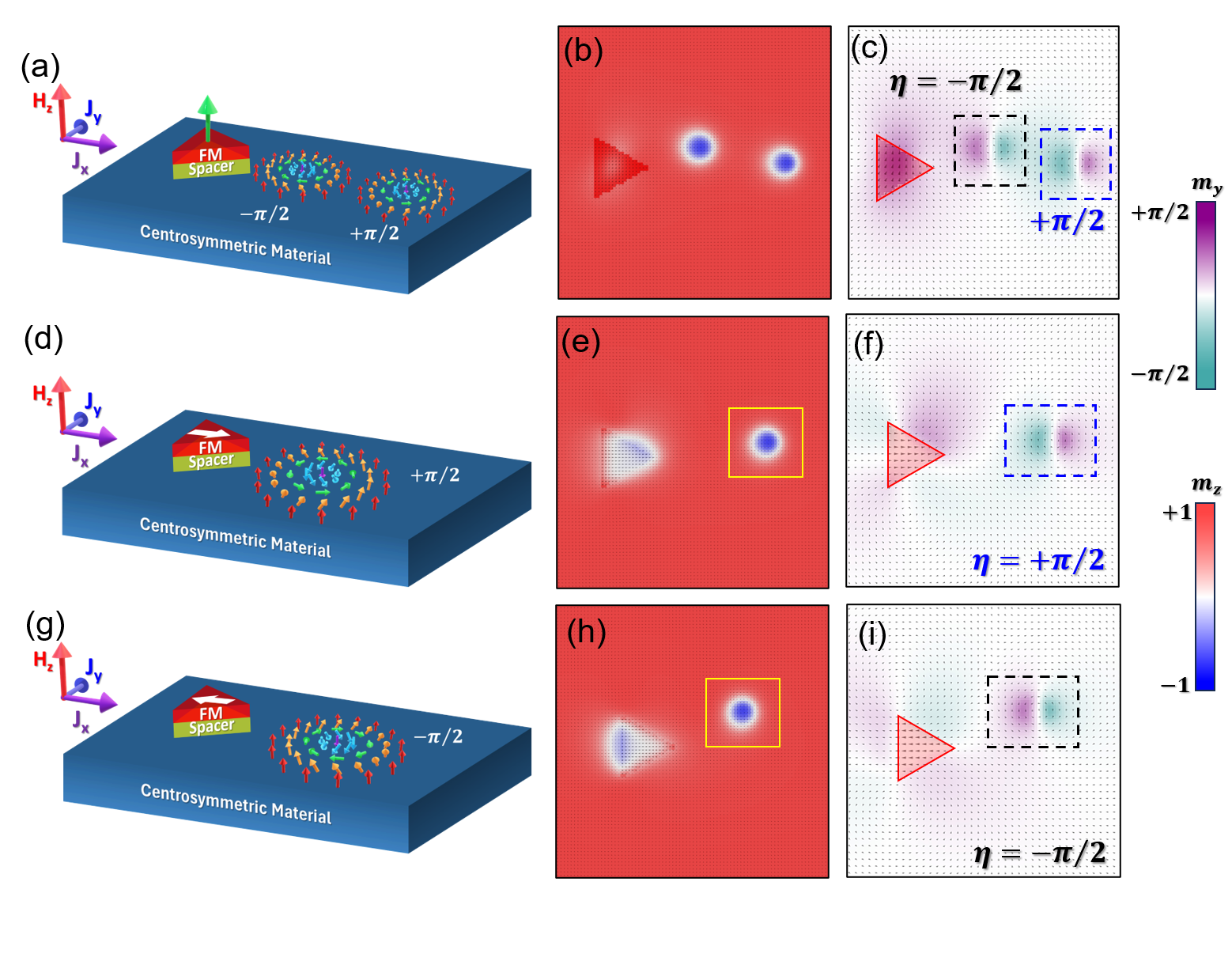}     \caption{\label{fig:FIG4}Deterministic and random skyrmion nucleation in FM/CM heterostructures under applied in-plane currents (J\textsubscript{x}, J\textsubscript{y}). (a) Schematic of helicity-locked skyrmion pair/random nucleation. Magnetization maps (m\textsubscript{y}, m\textsubscript{z}) show nucleation of skyrmions with helicities (b–c) The OP magnetization of the FM layer lifts the helicity degeneracy, enabling deterministic nucleation of CW/CCW skyrmion pairs, (d–f) Nucleation of CCW ($\eta$ = $+\pi/2$) skyrmions with FM magnetization along $+x$, (g–i) Nucleation of CW ($\eta$ = $-\pi/2$) skyrmions with FM magnetization reversed to $−x$. Color maps show the IP (m\textsubscript{y}) and OP (m\textsubscript{z}) magnetization components. 
	}
\end{figure*}

Using FM stripe with IP magnetization on the CM magnet with OP anisotropy, we demonstrate that the dipolar field acts as an effective symmetry-breaking interaction, selectively stabilizing one helicity over the other.
Next, we focus on using this concept to deterministically control the skyrmion helicity under applied in-plane currents (J\textsubscript{x}, J\textsubscript{y}), as illustrated in Fig. 4. 
Prior to applying the current pulse, the system was relaxed by a sequence of OP magnetic fields: an initial 300 mT field saturates the magnetization, followed by a reduction to 120 mT to reach equilibrium. 
For simulation, two pulse regimes are applied, a strong nucleation pulse at 12 × 10\textsuperscript{12} A/m\textsuperscript{2} followed by a significantly weaker driving pulse at 3 × 10\textsuperscript{12} A/m\textsuperscript{2}. 
In contrast, a short nucleation pulse produces skyrmions with random helicities, leading to stochastic spatial distributions without distinct helicity patterns as shown in \textbf{Supplement Video 2.}
In the triangular notch-like structure shown in Fig. 4a, the OP polarization of the FM layer modifies the energy landscape of the CM film, enabling deterministic formation of pair of skyrmions with CW and CCW helicity. %
Figure 4(b-c) shows magnetization maps of the skyrmion pairs with helicities locked by the FM layer, where the OP polarization of the FM breaks the intrinsic degeneracy of Bloch-type skyrmions in the CM layer. 
Symmetric, reproducible spatial patterns of these skyrmion pairs adjacent to the FM layer provide evidence for deterministic nucleation, which is shown in \textbf{Supporting Video 3}. 
When we take the FM layer polarized along $+x$ which imposes a preferred in-plane orientation on the Bloch-line spins [Fig. 4d], application of an in-plane current (J\textsubscript{x} = J\textsubscript{y}= 12 × 10\textsuperscript{12}\,A/m\textsuperscript{2}) to the CM layer results in nucleation of a skyrmion in the low-anisotropy FM region.
During the nucleation pulse, the spins in the Bloch line align along the $+x$ direction and undergo a CCW rotation driven by the polarization imposed by the FM layer. 
Upon applying a concurrent driving pulse, the bubble expansion is immediately suppressed, and the skyrmion detaches from the FM nucleation site, enabling controlled helicity-locked skyrmion generation. 
Figures 4(e-f) clearly show the nucleation of a skyrmion, while the helicity-resolved in-plane magnetization (m\textsubscript{y}) confirms its counterclockwise helicity ($\eta$ = $+\pi/2$). 
Under identical conditions, but with an extended current pulse width and the FM spins polarized along the $-x$ direction, as shown in Fig. 4g, the system stabilizes the opposite helicity ($\eta$ = $-\pi/2$), corresponding to a clockwise skyrmion. The nucleation and dynamics of the CW and CCW Bloch skyrmions are presented in \textbf{Supplementary Videos 4 and 5}, respectively. 
This observation is further supported by the helicity-resolved magnetization maps [Fig. 4(h–i)], which reveals an effective symmetry-breaking mechanism responsible for skyrmion nucleation in centrosymmetric systems.

\begin{figure*}[!t]
    \centering \includegraphics[width=0.9\textwidth,keepaspectratio]{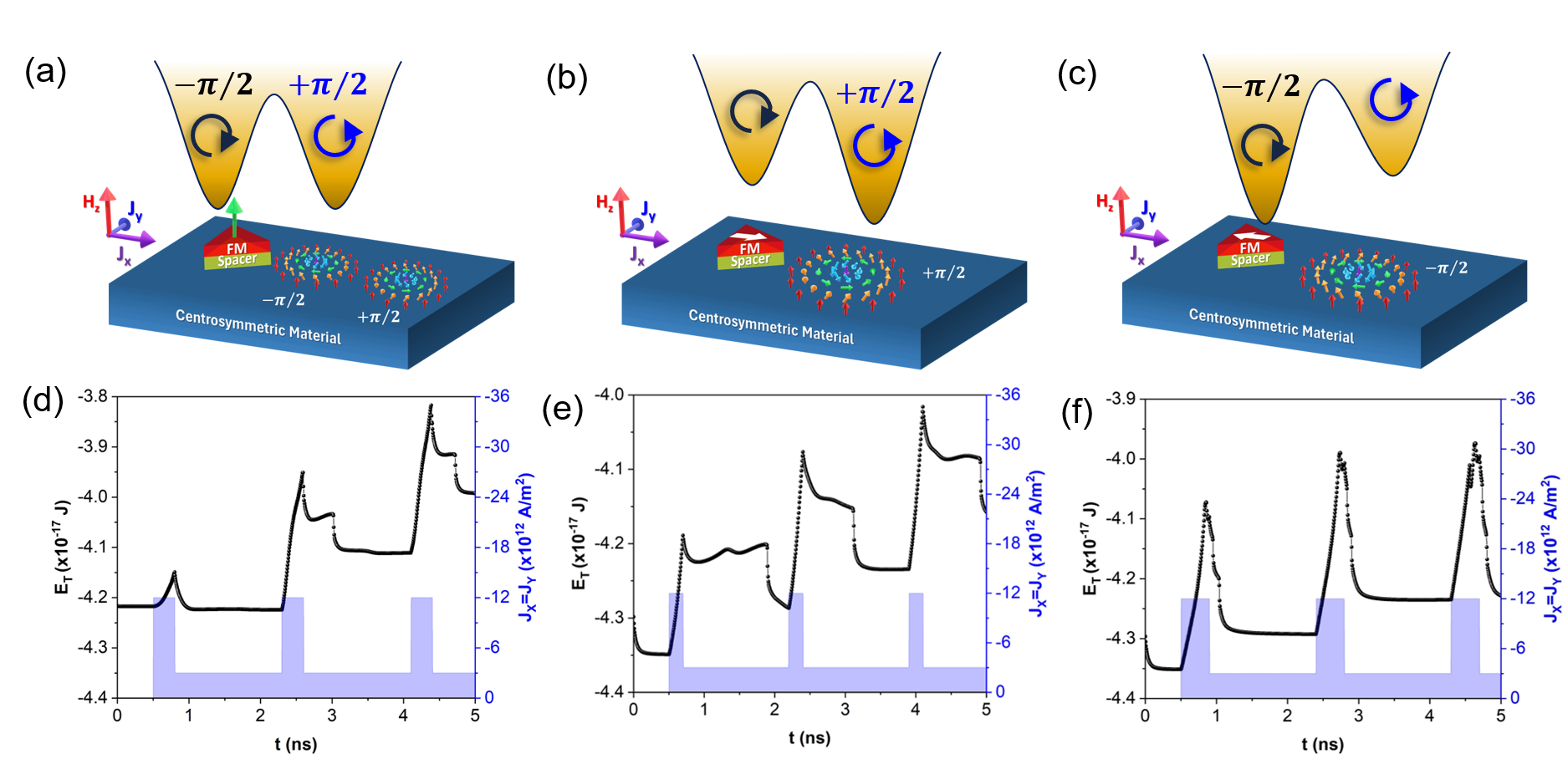}     \caption{\label{fig:FIG5}Energy landscape and helicity selection mechanism in FM/CM heterostructures (a) Schematic double-well potential illustrating helicity degeneracy ($\eta$ = $\pm\pi/2$) in a centrosymmetric film, (b–c) The FM layer introduces an effective dipolar interaction , where $\pm$ corresponds to the magnetization direction of the FM layer along $+x$, and $-x$ axis, respectively, lifting the helicity degeneracy and stabilizing a preferred helicity. (d–f) Time evolution of total system energy E\textsubscript{T} during current-induced skyrmion nucleation for FM polarized along $+z$ (CW-CCW pair), $+x$ (CCW), and $-x$ (CW), respectively.}
\end{figure*}

In centrosymmetric films, the helicity degeneracy states of skyrmions can be represented by a double-well potential, where skyrmions with $\eta$ = $+\pi/2$ and $-\pi/2$ states correspond to equal energy minima [Fig. 5a] \cite{nagaosa_topological_2013}. 
The introduction of an FM over-layer modifies this energy landscape by generating dipolar stray fields that couple to the spin texture of the centrosymmetric film. 
Depending on the orientation of the FM magnetization ($+x$ or $-x$), the effective dipolar interaction lifts the degeneracy of the helicity and stabilizes the preferred helicity state, as illustrated Fig. 5(b–c). 
This FM-mediated dipolar coupling acts as an effective symmetry-breaking interaction, analogous to the role played by DMI in non-centrosymmetric systems \cite{legrand_room-temperature_2020}, thereby enabling deterministic helicity selection. 
The time evolution of the total energy E\textsubscript{T} under in-plane current excitation (i.e., J\textsubscript{x} = J\textsubscript{y}) provides further evidence of the present phenomena [Figs. 5(d–f)]. 
When the FM stripe is magnetized along $+z$, the induced dipolar field does not preferentially break the helicity symmetry. 
The time evolution of E\textsubscript{T} exhibits multiple energy spikes corresponding to current-induced skyrmion nucleation events, followed by relaxation into stable pairs of CW-CCW helicity of nearly identical energy [Fig. 5d]. 
This symmetric energy profile highlights the pair-wise helicity nucleation characteristic of the magnetization configuration $+z$, consistent with the schematic double-well potential in Fig. 5a. 
The first nucleation pulse is insufficient to form a stable skyrmion but induces local spin reorientation through spin transfer torque (STT), effectively lowering the energy barrier for subsequent pulses and facilitating CW-CCW pair formation in later cycles. 
Random nucleation of the CW-CCW exhibits a profile of total system energy (E\textsubscript{T}) similar to that of nucleation of the CW-CCW pair, indicating comparable transient energy dynamics during the early stages of nucleation. 
However, deterministic pair nucleation requires a slightly longer nucleation pulse duration (approximately 0.05 ns more) than in the random case (see Supporting Fig. S4 \cite{SM}). 
The extended pulse allows sufficient STT and local spin relaxation, ensuring synchronized CW-CCW pair formation, whereas shorter pulses favor stochastic helicity selection and random skyrmion nucleation. 
This observation highlights the crucial role of pulse width in controlling the transition from random to deterministic helicity-locked skyrmion formation. 
Figure 5e illustrates the magnetized FM along the $+x$, where dipolar interactions induced by the FM layer create an energy landscape that lowers the nucleation barrier for CCW helicity skyrmions.
Upon pulsed current injection,E\textsubscript{T} exhibits transient increases corresponding to skyrmion nucleation, followed by rapid relaxation to lower energy states, demonstrating efficient nucleation with relatively short pulse widths due to the reduced energy barrier. 
In the case of the FM stripe polarized along the $-x$ that  promotes CW helicity skyrmions, the current pulses cause E\textsubscript{T} to increase with a prolonged nucleation dynamics and higher energy barriers that demand longer pulses for successful skyrmion formation [Fig. 5f]. 
This asymmetric nucleation behavior results directly from the asymmetric energy landscape created by the dipolar interaction of the FM. 
Consequently, this mechanism enables deterministic helicity control through FM orientation, allowing selective nucleation of CCW or CW skyrmions with tailored energy and pulse requirements. 
Collectively, these findings establish a clear correlation between interfacial exchange coupling, dipolar interaction, and current-induced dynamics in controlling skyrmion helicity and motion. 
The demonstrated FM-assisted helicity locking, drift-free transport, and deterministic nucleation provide a unified framework for reliable skyrmion manipulation in centrosymmetric systems. 
This integrated approach lays the groundwork for the development of DMI-independent, energy-efficient spintronic devices based on topologically protected spin textures.
\section{CONCLUSIONS}
In summary, we have shown through micromagnetic simulations that interfacing centrosymmetric magnetic (CM) thin films with ferromagnetic (FM) over-layers enables deterministic control of the helicity of Bloch-type skyrmions in the absence of DMI. The FM layers introduce dipolar and interfacial exchange fields that selectively stabilize either clockwise (CW) or counterclockwise (CCW) helicities, depending on the orientation of their in-plane magnetization. By systematically varying the FM geometry and interfacial coupling strength, we identify a transition from helicity-degenerate to helicity-locked regimes. Under strong coupling, the skyrmions become fully helicity-locked, exhibiting enhanced stability, improved coherence, and more efficient current-driven dynamics, including higher velocities and suppressed stochastic helicity switching. We further demonstrate robust FM-assisted nucleation schemes that reproducibly generate helicity-locked skyrmions and skyrmion pairs, overcoming the inherent randomness of helicity selection in centrosymmetric systems. Overall, these results establish FM/CM heterostructures as a powerful and flexible platform for engineering topological spin textures and advancing skyrmion-based technologies for memory, logic, and neuromorphic computing.

\begin{acknowledgments}
J.D. gratefully acknowledges the Department of Science and Technology – Anusandhan National Research Foundation (DST-ANRF), Government of India, for the National Post Doctoral Fellowship (N-PDF) and financial support (Fellowship Ref. No. PDF/2022/002556). AKN acknowledges the National Institute of Science Education and Research (NISER) and Department of Atomic Energy (DAE) for financial support.
\end{acknowledgments}

\section*{DATA AVAILABILITY}
The data supporting the findings of this article are not publicly available. The data are available from the authors upon reasonable request.
\bibliographystyle{apsrev4-2}
\bibliography{REF1}
\end{document}


\title{\textbf{Deterministic Helicity Locking of Bloch Skyrmions in Centrosymmetric Systems}}
\author{Jayaseelan Dhakshinamoorthy }
 \email{djn@niser.ac.in}
\author{Hitesh Chhabra}%
\author{Ajaya K Nayak}%
\email{ajaya@niser.ac.in}
\affiliation{%
School of Physical Sciences, National Institute of Science Education and Research, An OCC of Homi Bhabha National Institute, Jatni 752050, India\\
}%
\date{\today}
\maketitle
Simulations were initialized with random CM-layer magnetization and uniform FM-strip magnetization for conceptual illustration. A perpendicular external magnetic field of 120 mT was used to relax the system to equilibrium. Following relaxation, spin-polarized current pulses with density $J$ = ($x$, $y$, 0) and polarization $P$ = 0.6 were applied to induce skyrmion nucleation and motion. The current density vector $J $ was restricted to the IP directions ($x$, $y$), consistent with the Zhang–Li STT model \cite{albert_effect_2020}, which is well suited for thin-film geometries where lateral currents govern skyrmion dynamics. Magnetization dynamics was simulated using the Landau–Lifshitz–Gilbert (LLG) equation augmented with both adiabatic and non-adiabatic STT terms according to the Zhang–Li model \cite{hou_currentinduced_2023}:

\begin{equation}
\frac{d\mathbf{M}}{dt}
= -\gamma_0\,\mathbf{M}\times\mathbf{H}_{\mathrm{eff}}
+ \frac{\alpha}{M_s}\,\mathbf{M}\times\!\left(\frac{d\mathbf{M}}{dt}\right)
+ \frac{u}{M_s^2}\,\mathbf{M}\times\!\left(\frac{\partial\mathbf{M}}{\partial x}\times\mathbf{M}\right)
- \frac{\beta\,u}{M_s}\,\mathbf{M}\times\!\left(\frac{\partial\mathbf{M}}{\partial x}\right)
\end{equation}

where $\mathbf{H}_{\mathrm{eff}} = -\frac{1}{\mu_0}\frac{\delta E}{\delta \mathbf{M}}$, is the effective magnetic field and the parameter '$u$' quantifies the STT strength, $u = \left| \frac{\gamma_0 \hbar}{\mu_0 e} \right| \frac{J P}{2 M_s}$. Here $\mu_0$ is the vacuum permeability, $\gamma_0 $ is the gyromagnetic ratio, $\alpha$ is the Gilbert damping constant, $ħ$ is the reduced Planck constant, $e$ is the electron charge, $J=|J|$ is the applied current density and $P$ is the spin polarization. The dimensionless constant $\beta$ characterizes the strength of the non-adiabatic torque relative to the adiabatic contribution. Theoretical models and prior simulations consistently report non-adiabaticity values in the range of 0.02 to 0.1\cite{hou_currentinduced_2023, bernstein_spin-torque_2025}, making  $\beta$ = 0.05 a reasonable choice for modeling skyrmion dynamics under STT. To facilitate practical implementation, we slightly adjusted the $M_s$, $A_{\text{ex}}^*$ and $K_u$ values of the FM and CM layers, and simulated the magnetization states of oppositely oriented FM strips under both in-plane and OP magnetic fields, as illustrated in \textbf{Supplementary Video 1}. Nevertheless, to demonstrate the concept of helicity modulation and Bloch skyrmion nucleation, the previously specified parameters were consistently used throughout this study. Numerical checks with different mesh sizes confirmed that the observed effects were not artifacts of the grid. The larger anisotropy in the FM regions ensured stable magnetization during current application, effectively replicating experimental FM behavior.
\begin{figure*}[t]
    \centering \includegraphics[width=0.9\textwidth,keepaspectratio]{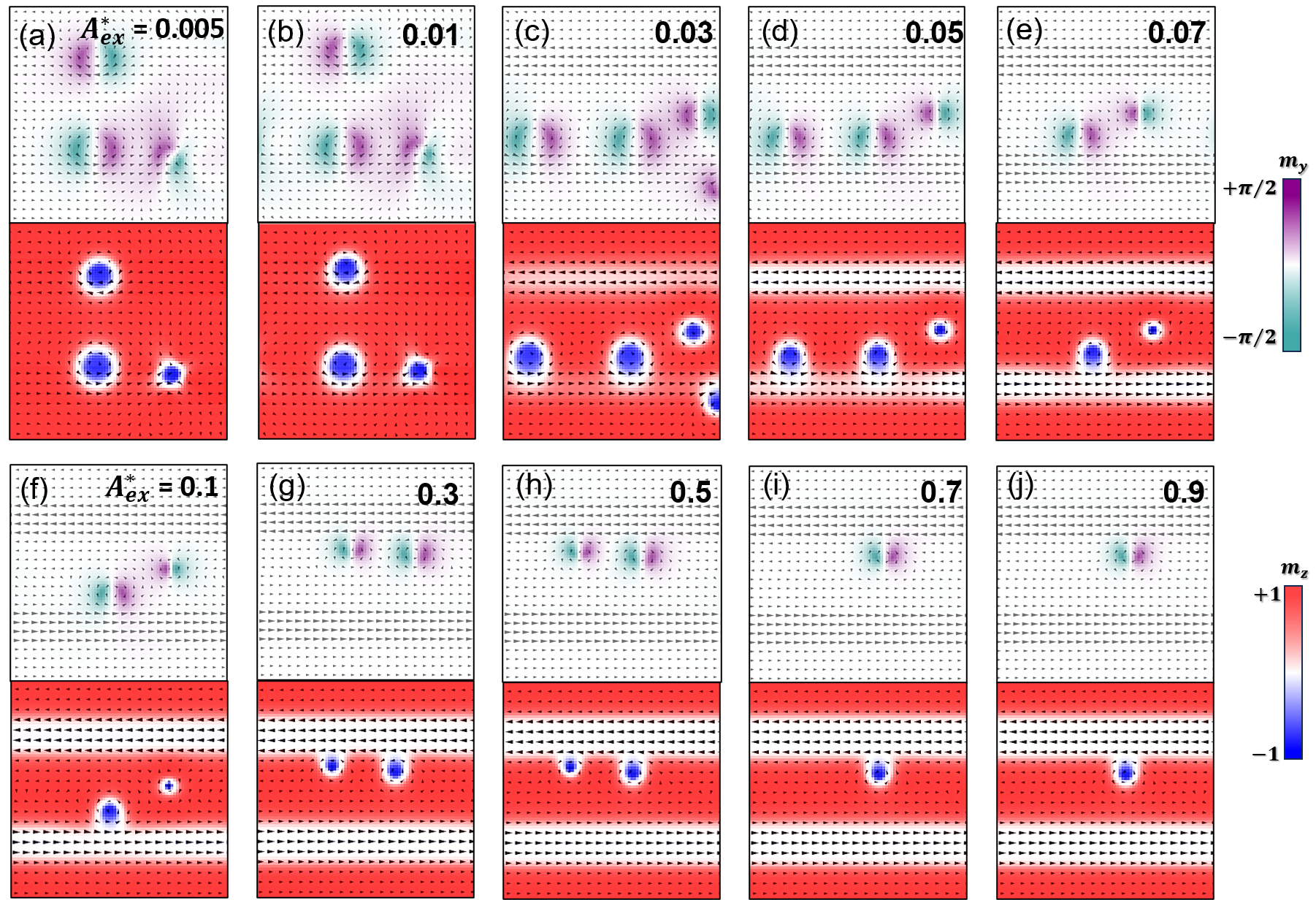} 
    \caption{\label{fig:widepng}Micromagnetic map showing the effect of reduced interfacial exchange scale Panels (a–f) correspond to very weak coupling , where skyrmions with both helicities coexist, and panels (g–j) represent stronger coupling , where only isolated helicity-locked Bloch skyrmions ($\eta$ = $+\pi/2$) remain stable. Color maps indicate m\textsubscript{y} (IP) and m\textsubscript{z} (OP) components.}
\end{figure*}
\textbf{Figure S1}. illustrating the role of reduced interfacial exchange coupling $A_{\text{ex}}^*$ on skyrmion stability and helicity selection. The parameter $A_{\text{ex}}^*$  represents the normalized exchange scaling between the FM and CM layers. In very weak coupling ($A_{\text{ex}}^*$=0.005 - 0.1, \textbf{Fig. S1(a–f)),} the dipolar interaction of FM strips is insufficient to completely remove the degeneracy of helicity. As a result, Bloch skyrmions with both helicities $ \eta = \pm \pi/2 $ coexist with mixed states, and anti-skyrmions occasionally appear. The corresponding in-plane helicity-resolved my magnetization maps (upper panels) reveal multiple spin textures with distinct helicities, consistent with the metastability of degenerate states in the weak-coupling regime. As $A_{\text{ex}}^*$ increases, (0.3 to 0.9, \textbf{Fig. S1(g–j))}, the interlayer exchange and the dipolar coupling from the FM become stronger, progressively stabilizing a unique helicity state. In this regime, only isolated helicity-locked Bloch skyrmions with $ \eta = +\pi/2 $ persist, while $ \eta = -\pi/2 $ skyrmions and anti-skyrmions are eliminated. This transition indicates that interfacial exchange scaling serves as a control knob for deterministic helicity selection in centrosymmetric systems. The bottom panels show the magnetization $m_z$, where isolated skyrmion cores are clearly resolved. The helicity-resolved in-plane distributions highlight the evolution from helicity-degenerate textures at weak coupling to helicity-locked states at strong coupling.
\begin{figure}[t]
    \centering \includegraphics[width=0.5\textwidth,keepaspectratio]{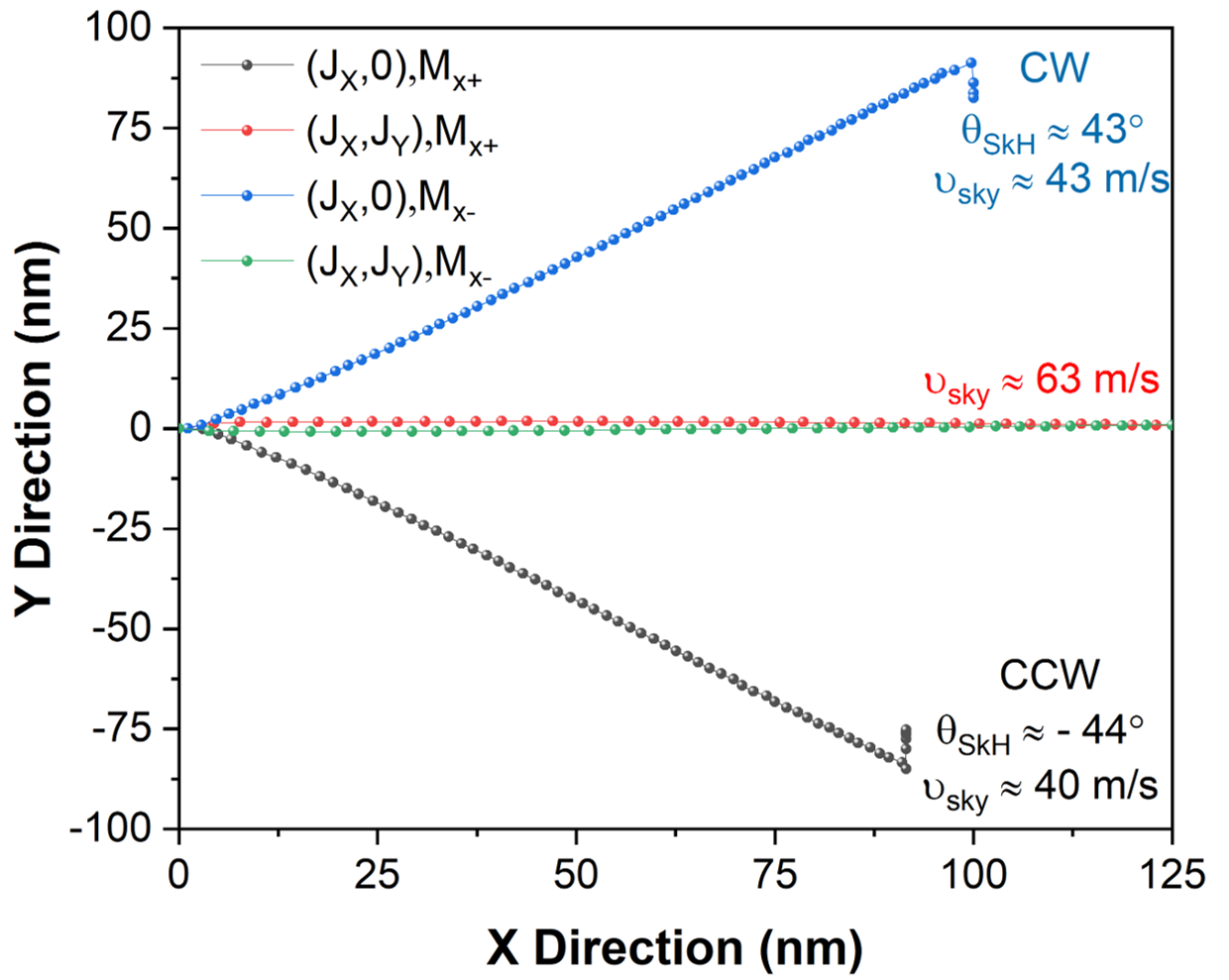} 
    \caption{\label{fig:widepng}Current-driven skyrmion motion under different magnetization and current configurations. }
\end{figure}
\textbf{Figure S2} illustrates the simulated motion of skyrmions driven by IP currents under different magnetization orientations and current configurations. The simulations were performed with a Gilbert damping constant of $\alpha$ = 0.9 and a non-adiabaticity parameter of $\beta$ = 0.05. A large $\alpha$ suppresses magnetization precession and strongly damps gyrotropic oscillations, resulting in smooth, monotonic skyrmion motion. This ensures that the skyrmion follows a stable trajectory along the current direction without random fluctuations or overshoot. The higher $\alpha$ effectively reduces the magnitude of the skyrmion Hall angle ($\theta_\text{SkH}$), since the damping term counteracts the transverse gyrotropic force responsible for the Hall deflection. Consequently, in \textbf{Figure S2}, the skyrmions move steadily along well-defined paths with moderate Hall angles, and their motion remains deterministic rather than stochastic. For the configuration ($J_x$, 0), $M_x^+$ (black curve), the skyrmion travels mainly along the $−y$ direction with a velocity $\nu_\text{sky}$ $\approx$  40 m/s. Reversing the background magnetization to $M_x^-$(blue curve) reverses the gyrovector polarity, yielding a right$-$handed (CW) motion with $\nu_\text{sky}\approx 43$ m/s, and $\theta_\text{SkH} \approx 43^o$. The opposite signs of the transverse velocity component ($\nu_y$) reflect chirality reversal (CW $\leftrightarrow$ CCW) arising from the change in gyrovector orientation. The slightly lower speed observed in the CW case is due to enhanced damping and reduced effective STT projection. When both in-plane current components ($J_x$, $J_y$) are applied (red and green curves), the transverse velocity nearly vanishes ($\theta_\text{SkH}$ $\approx$  $0^o$) due to vector cancellation of the gyrotropic term, resulting in straight-line motion. Such suppression of the skyrmion Hall effect enables drift-free trajectories desirable for racetrack-type skyrmion devices.
\begin{figure*}[t]
    \centering \includegraphics[width=0.9\textwidth,keepaspectratio]{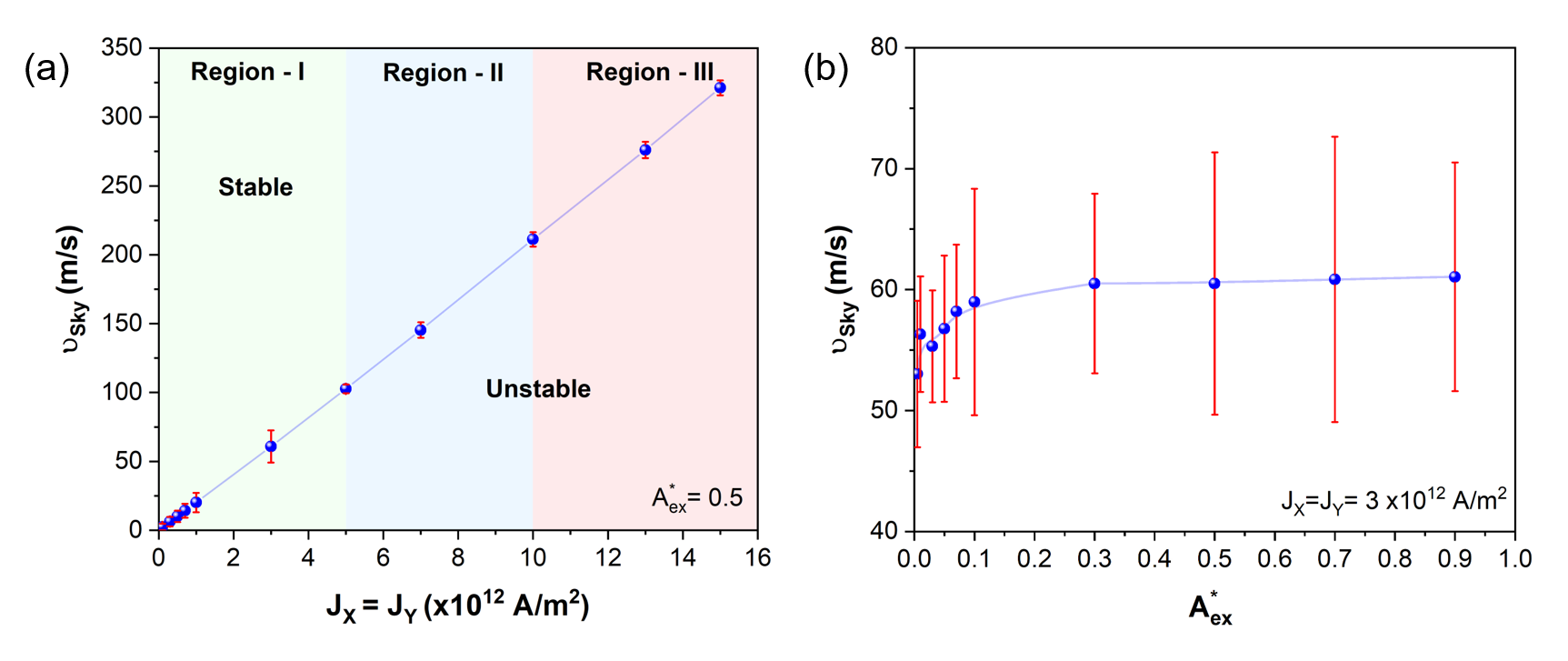} 
\caption{\label{fig:widepng}Current-driven skyrmion dynamics in FM/CM heterostructures. (a) Skyrmion velocity ($\nu_\text{sky}$) as a function of in-plane current density ($J_x$ = $ J_y$), at $A_{\text{ex}}^*$ = 0.5, showing three dynamical regimes, (b) Skyrmion velocity ($\nu_\text{sky}$) versus interfacial exchange coupling $A_{\text{ex}}^*$ at a fixed current density $J_x$ = $ J_y$ = 3 × 10\textsuperscript{12}A/m\textsuperscript{2}  }
\end{figure*}

The influence of applied current density and interfacial exchange coupling on skyrmion dynamics in FM/CM heterostructures is shown in \textbf{Fig. S3}. At a fixed reduced interfacial exchange scale ($A_{\text{ex}}^*$  = 0.5), the skyrmion velocity ($\nu_\text{sky}$) increases linearly with the applied IP current density ($J_x$ = $J_y$), up to 15 × 10\textsuperscript{12}  A/m\textsuperscript{2} [\textbf{Fig. S3(a)}]. The velocity–current characteristics exhibit three distinct regimes: (i) Region I (low-current regime), where skyrmions move slowly while remaining stable, dominated by the balance between STT and damping; (ii) Region II (intermediate current regime), where the velocity increases sharply with minimal distortion of the spin texture; and (iii) Region III (high-current regime), where skyrmions become unstable, exhibiting deformation and velocity fluctuations consistent with earlier reports on current-driven skyrmion transport \cite{litzius_skyrmion_2017}. At a fixed current density ($J_x$ = $J_y$ = 3 × 10\textsuperscript{12} A/m\textsuperscript{2} ), the skyrmion velocity strongly depends on the interfacial exchange coupling [\textbf{Fig. S3(b)}]. Specifically, $\nu_\text{sky}$ increases rapidly with small increases in $A_{\text{ex}}^*$  and then saturates beyond $A_{\text{ex}}^*$  $\geq$ 0.3. At weak interfacial exchange ($A_{\text{ex}}^*$$< $ 0.3), the FM and CM layers are only weakly coupled, and the degeneracy of Bloch skyrmions ($ \eta = \pm \pi/2 $ ) is not fully lifted. This results in metastable helicity-degenerate states that introduce fluctuations in skyrmion dynamics, thereby reducing the effective velocity. As $A_{\text{ex}}^*$  increases slightly, the exchange field from the FM begins to polarize the Bloch-line, progressively enforcing helicity locking ($ \eta = +\pi/2 $ ). This suppresses stochastic helicity switching and deformation, allowing skyrmions to respond more coherently to applied currents, which leads to a rapid rise in velocity. For $A_{\text{ex}}^*$ $\geq$ 0.3, the FM coupling is sufficiently strong to completely lift the helicity degeneracy, stabilizing a unique helicity state. In this regime, further increases in $A_{\text{ex}}^*$  do not provide additional stabilization, and the skyrmion dynamics reach an optimal coherent state where the velocity saturates. This transition demonstrates that interfacial exchange coupling not only fixes skyrmion helicity, but also enhances their dynamical stability and transport efficiency.

\begin{figure*}[t]
    \centering \includegraphics[width=0.9\textwidth,keepaspectratio]{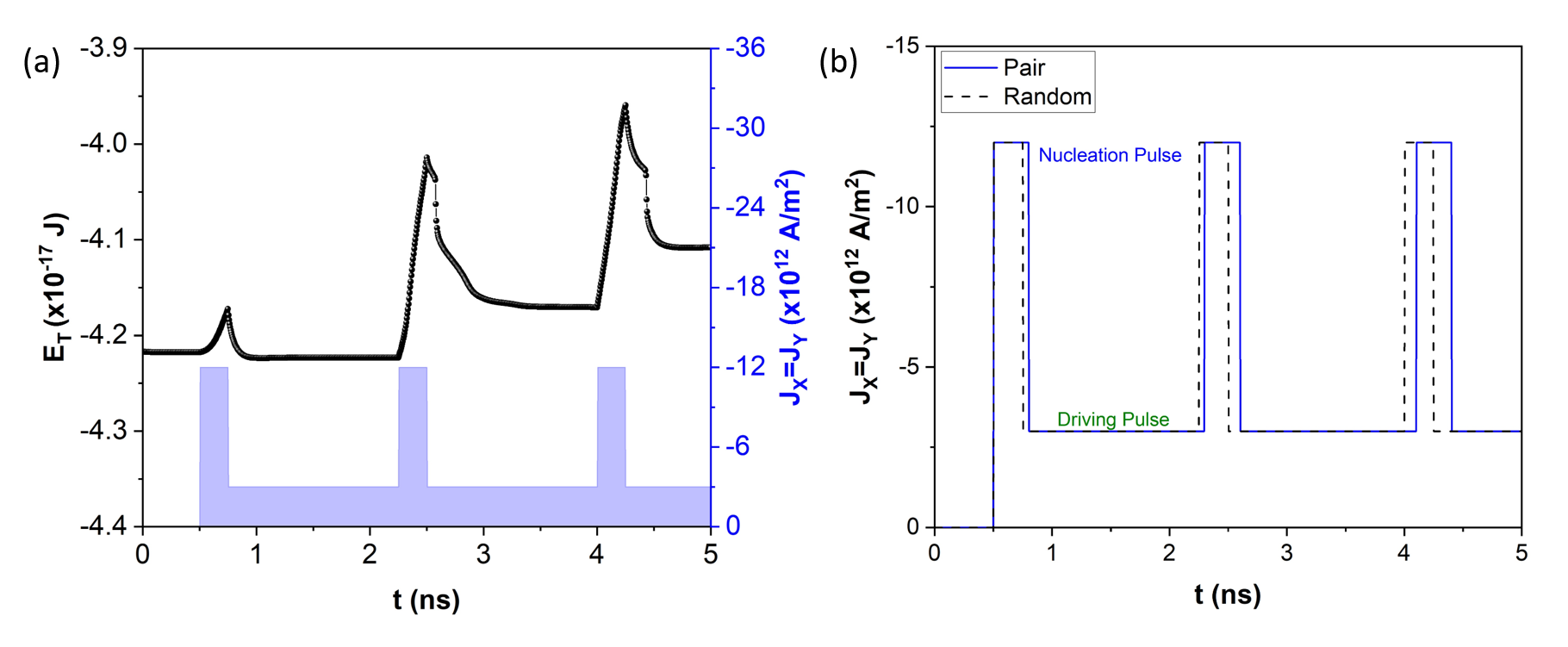} 
\caption{\label{fig:widepng}(a) Time evolution of total system energy E\textsubscript{T} during current-induced skyrmion nucleation for FM polarized along $+z$ (CW-CCW random favored), (b) Time sequence of applied current pulses: a nucleation pulse (12 x 10\textsuperscript{12} A/m\textsuperscript{2}) followed by a driving pulse (3 x 10\textsuperscript{12} A/m\textsuperscript{2}), demonstrating deterministic pair creation (solid line) compared with random nucleation (dotted line). }
\end{figure*}

\textbf{Figure S4 }presents the time evolution of the total energy of the system ($E_T$) during the nucleation of the skyrmion induced by current and the corresponding time sequence of applied current pulses used to compare the formation of a deterministic CW–CCW pair with random nucleation. Simulations were performed for the FM/CM heterostructure with the FM magnetized along $+z$, where symmetric dipolar fields promote the generation of skyrmion pairs of helicity. As shown in \textbf{Fig. S4a,} both random nucleation and pair nucleation exhibit similar oscillatory profiles in $E_T$, characterized by discrete energy spikes corresponding to successive nucleation events followed by rapid relaxation. The first pulse causes a slight rise in $E_T$ but does not form a stable skyrmion; instead, it induces local spin reorientation via STT, lowering the effective energy barrier for subsequent nucleation. Subsequent pulses lead to higher and broader energy peaks associated with the formation of stable CW–CCW skyrmion pairs. This cumulative energy evolution indicates that early current pulses precondition the magnetic texture, enabling efficient nucleation in subsequent cycles. \textbf{Figure S4b} shows the corresponding pulse profiles used for deterministic and random nucleation. A strong nucleation pulse (12×10\textsuperscript{12}A/m\textsuperscript{2} ) is followed by a weaker driving pulse (3×10\textsuperscript{12}A/m\textsuperscript{2} ) to stabilize and drive the generated skyrmions. Deterministic nucleation of the CW-CCW pair requires a slightly longer nucleation pulse duration (approximately 0.05 ns more) than random nucleation. The extended pulse provides additional STT energy and relaxation time, ensuring symmetric helicity formation, whereas shorter pulses (0.25 ns) favor stochastic helicity selection, leading to random skyrmion configurations. These results confirm that precise control over pulse width and sequence is critical for reproducibly forming helicity-locked skyrmion pairs. Longer nucleation pulses promote deterministic generation of the CW-CCW pair, while shorter pulses produce uncorrelated random helicity states. This finding underscores the importance of current-pulse engineering in achieving reliable, energy-efficient skyrmion nucleation in centrosymmetric FM/CM heterostructures.
 
             \bibliography{REF1}